\documentclass{article}
\usepackage{ijcai25}

\usepackage{times}
\usepackage{natbib}
\usepackage{soul}
\usepackage{url}
\usepackage[hidelinks]{hyperref}
\usepackage[utf8]{inputenc}
\usepackage[small]{caption}
\usepackage{graphicx}
\usepackage{amsmath}
\usepackage{amsthm}
\usepackage{booktabs}
\usepackage{algorithm}
\usepackage{algorithmic}
\usepackage[switch]{lineno}
\usepackage{colortbl} 
\usepackage{tikz}
\usetikzlibrary{bayesnet}
\usepackage{algorithm}
\usepackage{algorithmic}

\usepackage{fancyhdr}

\title{Building Trust in Illiquid Markets: an AI-Powered Replication of Private Equity Funds}

\author{
E. Benhamou$^{1,2}$ \and
JJ. Ohana$^1$ \and
B. Guez$^1$ \and
E. Setrouk$^1$ \and
T. Jacquot$^1$ \\
\affiliations
$^1$Ai for Alpha \quad
$^2$Dauphine PSL \quad
\emails
\{first\_name.last\_name\}@aiforalpha.com
}

\begin{document}
\maketitle

\begin{abstract}
In response to growing demand for resilient and transparent financial instruments, we introduce a novel framework for replicating private equity (PE) performance using liquid, AI-enhanced strategies. Despite historically delivering robust returns, private equity's inherent illiquidity and lack of transparency raise significant concerns regarding investor trust and systemic stability, particularly in periods of heightened market volatility. Our method uses advanced graphical models to decode liquid PE proxies and incorporates asymmetric risk adjustments that emulate private equity's unique performance dynamics. The result is a liquid, scalable solution that aligns closely with traditional quarterly PE benchmarks like Cambridge Associates and Preqin. This approach enhances portfolio resilience and contributes to the ongoing discourse on safe asset innovation, supporting market stability and investor confidence.
\end{abstract}

\section{Introduction}\label{sec:Introduction}
Trust in financial markets is increasingly tied to the ability of institutions to manage risk transparently, especially in illiquid asset classes. Among these, private equity (PE) has emerged over the past two decades as a dominant source of long-term returns. As highlighted by leading consultancy reports \cite{bcg2020global, mckinsey2020annual, McKinsey2023}, global assets under management (AUM) in private equity exceeded \$13 trillion as of June 2023, growing steadily at nearly 2\% annually since 2018. Major institutional investors have responded by heavily allocating capital to this space: CalPERS recently raised its private market target allocation from 33\% to 40\%, contributing over \$15 billion to co-investments in just 18 months. Other major players such as Singapore’s GIC, Harvard University, CPPIB, and Yale have similarly committed between 32\% and 39\% of their portfolios to private markets.

While the long-term return potential of private equity is undisputed, its illiquidity raises systemic concerns. PE investments require extended capital lock-up periods, lack standardized pricing, and depend on exit timing that can be misaligned with investor liquidity needs. In periods of market stress or when institutional liabilities increase, these characteristics render private equity portfolios especially burdensome to rebalance or unwind. The inability to liquidate holdings at fair value undermines both portfolio resilience and investor confidence.

This illiquidity paradox—high-return potential at the cost of low liquidity—has led to significant interest in developing liquid alternatives that can replicate the economic exposures of PE without its structural constraints. However, as argued by \cite{Kaplan_2005, Ang2014, ang2014estimating, Franzoni2012}, private equity’s outperformance may be intrinsically linked to its illiquidity premium. Investors are often compensated for the lack of liquidity with higher returns, raising the question of whether these returns can truly be replicated using liquid instruments.

Nonetheless, academic and industry researchers have proposed various liquid proxies and replication strategies. \cite{Stafford2021} developed a method based on small public stocks with PE-like characteristics, while \cite{thomson_reuters_pe_index} introduced a sector-based approach for greater scalability. These strategies partially address the liquidity issue, but they struggle to capture essential features of PE value creation—such as operational improvements, governance changes, and the discretion over exit timing—emphasized in \cite{Kaplan2009leveraged}. Moreover, many of these proxies use assets that are themselves not very liquid or scalable.

A key insight emerging from recent literature is the importance of asymmetric return patterns in private equity replication. PE returns tend to be positively skewed, in part due to downside smoothing and the avoidance of forced exits in declining markets. Studies such as \cite{Ang2014} and \cite{Franzoni2012} demonstrate that incorporating these asymmetries—particularly those arising in risk-off environments—can enhance replication fidelity. Capturing this behavior is essential to producing not only accurate, but trustworthy and resilient, synthetic exposures.

This paper explores how advanced AI techniques and financial engineering can be used to create liquid instruments that approximate the performance of private equity funds while enhancing market transparency and stability. By doing so, we contribute to the ongoing discussion around constructing "safe assets" that meet the dual objectives of return generation and systemic resilience.

\subsection{Innovation}\label{sec:Innovation}
Our contributions to the literature on private equity replication are threefold:

\begin{itemize}
    \item \textbf{Liquidity Transformation:} We propose a strategy that uses highly liquid equity index futures to replicate the performance of private equity portfolios, thereby addressing institutional liquidity constraints without sacrificing scalability.
    \item \textbf{AI-Based Precision:} We apply advanced machine learning techniques, specifically graphical models, to decode the dynamics of liquid PE proxies. This enhances replication fidelity compared to traditional linear or Kalman-based models.
    \item \textbf{Asymmetric Risk Adjustment:} Our framework introduces asymmetry in the return modeling process to account for private equity’s reduced downside volatility, manager discretion in exit timing, and strategic value creation, resulting in a return profile that more closely mirrors actual PE funds.
\end{itemize}

\subsection{Structure of the Paper}
This paper is structured to address the central question of how AI can be leveraged to replicate private equity (PE) in a manner that enhances trust, transparency, and liquidity in financial markets—especially during periods of systemic stress.

Section~\ref{sec:Literature Review} reviews the foundational literature on private equity, with particular emphasis on institutional demand, the three pillars of PE value creation (corporate governance, operational optimization, and financial engineering), and the challenges that arise from the asset class’s inherent illiquidity. We also examine recent advances in liquid replication strategies and their implications for market trust and accessibility.

Section~\ref{sec:Methodology} introduces our AI-powered framework for replicating PE performance. We detail a two-step process that decodes liquid proxies using graphical models and incorporates asymmetric return dynamics to better match the behavior of traditional PE funds. This section also explains how our methodological choices directly address issues of scalability, transparency, and systemic resilience.

Section~\ref{sec:Results and statistics} presents empirical results demonstrating how our approach compares to traditional benchmarks. We assess performance metrics such as Sharpe ratio, drawdown, and correlation to evaluate the fidelity and robustness of the replication. Special attention is given to how our approach mitigates downside risk and enhances liquidity.

Finally, Section~\ref{sec:Conclusion} concludes with a summary of key contributions and discusses the broader implications of our findings for financial stability and safe asset design. We also outline future research directions, including the extension of our methodology to other illiquid asset classes and its potential role in macroprudential policy.

\section{Literature Review}\label{sec:Literature Review}

\subsection{Private Equity, Value Creation, and the Trust-Liquidity Trade-Off}
Private equity (PE) has long been recognized as a high-performing asset class due to its unique approach to value creation. Unlike venture capital, PE typically involves acquiring controlling stakes in mature companies, applying targeted strategies across three central pillars: corporate governance, operational optimization, and financial engineering \cite{Kaplan2009leveraged}. These mechanisms empower PE sponsors to introduce strong managerial incentives, streamline operations, and enhance capital efficiency.

Managerial alignment is particularly notable in private equity. As documented in \cite{Kaplan_2005} and \cite{acharya2009private}, transitioning from public to private ownership often results in executives acquiring significant equity stakes. This illiquid compensation structure aligns long-term interests and discourages short-term value manipulation, thereby contributing to sustained performance gains.

Financial engineering, another key component, involves optimizing capital structures and utilizing leverage to extract value. \cite{Jensen1986agency} highlights how debt imposes discipline by limiting excess free cash flows, though this benefit is tempered by the risk of overleveraging \cite{axelson2013borrow}. Meanwhile, operational improvements have become increasingly sophisticated since the 1990s, with top-performing PE firms bringing sector-specific knowledge and active portfolio support to enhance productivity and profitability \cite{Kaplan_2005, acharya2009private}.

These pillars of value creation help explain PE’s superior risk-return profile. However, this performance comes at the cost of liquidity and transparency. The long-term, illiquid nature of PE investments introduces significant challenges, especially in times of market stress when exit opportunities vanish and valuations become opaque. As \cite{doskeland2018evaluating} argue, this illiquidity necessitates a premium, particularly because it hinders portfolio rebalancing and liability management for institutions. Diversification across vintage years \cite{Robinson2016} helps, but the core trust challenge remains: how can institutions maintain exposure to PE-like returns without sacrificing liquidity and transparency?

\subsection{Replicating PE: Toward Liquid and Trustworthy Alternatives}
To address this trust-liquidity paradox, researchers have explored how public market instruments might replicate PE returns while preserving liquidity. These efforts align with the broader objective of designing financial products that promote transparency and systemic stability.

\cite{rasmussen2015leveraged} were early contributors to this field, developing a public equities-based approach using small, leveraged, undervalued stocks. Their portfolios—focused on factors such as debt reduction and asset turnover—achieved high alphas under both CAPM and Fama-French models. Notably, their results reflected the potential to reproduce aspects of PE returns using disciplined, rules-based public strategies. However, volatility remained elevated compared to private equity, revealing a core limitation: daily marked-to-market pricing cannot fully replicate the smoothed return profile of illiquid assets.

Building on this, \cite{Stafford2021} proposed a more structured replication using portfolios of small, low-multiple public companies with added leverage and extended holding periods. While this approach effectively mimics buyout fund exposures, it suffers from limitations in data quality, execution scalability, and the challenge of modeling discretionary capital deployment.

In parallel, the Thomson Reuters Private Equity Index \cite{thomson_reuters_pe_index} offered a sector-based proxy for PE performance, improving scalability by relying on industry-level trends rather than individual security selection. However, the loss of granularity may reduce its ability to capture the nuanced drivers of PE value creation—such as governance interventions or sales timing discretion \cite{Kaplan2009leveraged}.

\cite{gupta2021valuing} added further perspective by decomposing PE performance into cash flow components and constructing replicating portfolios from listed equity and fixed income exposures. Their "strip-by-strip" valuation framework sheds light on the macro and micro risk drivers of PE but stops short of producing daily liquid strategies that institutions can deploy at scale.

Despite these advances, a central challenge persists: how to engineer daily tradable instruments that are not only return-aligned with private equity but also exhibit lower drawdowns, smoother performance, and trustworthy signals in times of stress. Many replication models rely on backward-looking regressions or linear mappings, which fail to capture the nonlinear, asymmetric dynamics of private equity returns \cite{Franzoni2012, Ang2014}.

This paper builds on these efforts by introducing a framework that uses AI and graphical modeling to replicate PE exposures through liquid instruments while explicitly incorporating return asymmetries and performance smoothing. In doing so, we contribute to the design of transparent, resilient financial products—ones capable of serving as safe assets in portfolios that require both yield and trust.

\section{Methodology}\label{sec:Methodology}

\subsection{Goal}
This study aims to develop a daily liquid replication framework that can emulate the performance of leading private equity (PE) benchmarks such as Cambridge Associates, Preqin, and Bloomberg’s Private Equity Buyout and All Indexes. These benchmarks are characterized by superior Sharpe ratios (typically around 1.5 or higher), low drawdowns, and annualized returns between 11\% and 15\%. However, their quarterly reporting frequency and illiquid nature make them unsuitable for real-time portfolio implementation. This disconnect poses a challenge to institutional investors who seek the performance benefits of private equity while maintaining liquidity, transparency, and trust—especially during periods of systemic stress.

\subsection{Performance of Private Equity Benchmarks}

Tables~\ref{tab:quaterly benchmarks} and~\ref{tab:bloomberg} present a summary of key risk-return characteristics for established PE benchmarks. Notably, these benchmarks consistently outperform traditional equity indexes on a risk-adjusted basis. Their low volatility and drawdowns, along with high Sharpe and Sortino ratios, exemplify the traits of a "safe asset" class—yet one currently inaccessible for daily liquid replication.

While the Cambridge Associates benchmark shows relative independence, the Preqin, PEALL, and PEBUY benchmarks are closely correlated (Table~\ref{tab:quarterly benchmark correlations}), reinforcing the consistency of their risk-return profiles across methodologies.

\begin{table}[htbp]
  \centering
  \caption{Performance of Traditional Private Equity Indexes
  \label{tab:quaterly benchmarks}}
  \resizebox{\columnwidth}{!}{%
    \begin{tabular}{llr}
    \toprule
           & \textbf{Cambridge Associates (CA)} & \textbf{Preqin} \\
    \midrule
    Start Date & 31/03/2011 & 31/03/2011 \\
    End Date & 29/12/2023 & 29/12/2023 \\
    Annual Return & 13.9\% & 14.2\% \\
    Annual Volatility & 8.9\% & 7.5\% \\
    Skew   & -0.27  & 0.06 \\
    Kurtosis & 1.64   & 1.46 \\
    Sharpe Ratio & 1.56   & 1.89 \\
    Sortino Ratio & 2.18   & 2.66 \\
    Max DD & 9.5\% & 7.3\% \\
    10\% Worst DD & 3.7\% & 1.7\% \\
    Return/maxDD & 1.5    & 1.9 \\
    Return/Worst 10\% DD & 3.8    & 8.5 \\
    Sampling & quarterly & quarterly \\
    \bottomrule
    \end{tabular}%
  }
\end{table}

\begin{table}[htbp]
  \centering
  \caption{Correlation between quarterly benchmarks \label{tab:quarterly benchmark correlations}}
  \resizebox{0.75 \columnwidth}{!}{%
    \begin{tabular}{|c|rrrr}
    \toprule
           & \multicolumn{1}{c}{\textbf{CA}} & \multicolumn{1}{c}{\textbf{Prequin}} & \multicolumn{1}{c}{\textbf{PEBUY}} & \multicolumn{1}{c}{\textbf{PEALL}} \\
    \midrule
    \textbf{CA} & 100\%  &        &        &  \\
    \textbf{Prequin} & 61\%   & 100\%  &        &  \\
    \textbf{PEBUY} & 75\%   & 91\%   & 100\%  &  \\
    \textbf{PEALL} & 70\%   & 96\%   & 97\%   & 100\% \\
    \cmidrule{1-1}    
  \end{tabular}
  }
\end{table}%

\subsection{Challenges and Alternatives with Daily Indexes}

Efforts to develop daily proxies for private equity have led to the emergence of indices such as Erik Stafford’s SHPEISM, the Thomson Reuters Private Equity Benchmark, and the S\&P Listed Private Equity Index. These proxies offer daily quotations but suffer from substantially lower Sharpe ratios (below 0.55) and extreme drawdowns exceeding 40\% (Table~\ref{tab:daily_index}). Their return profiles are noisier and more vulnerable to tail events—undermining trust and their viability as substitutes for quarterly PE benchmarks in institutional settings.

\begin{table}[htbp]
  \centering
  \caption{Performance of Daily Liquid Indexes \label{tab:daily_index}}
  \resizebox{\columnwidth}{!}{%
    \begin{tabular}{lrrr}
    \toprule
           & \textbf{Stafford} & \textbf{TR}     & \textbf{Listed PE} \\
    \midrule
    \midrule
    Start Date & 31/03/2011 & 31/03/2011 & 31/03/2011 \\
    End Date & 21/01/2025 & 21/01/2025 & 21/01/2025 \\
    Annual Return & 10.90\% & 12.50\% & 10.90\% \\
    Annual Volatility & 25.90\% & 24.80\% & 20.20\% \\
    Skew   & -0.33  & -0.64  & -0.78 \\
    Kurtosis & 3.02   & 1.55   & 18.2 \\
    Sharpe Ratio & 0.42   & 0.5    & 0.54 \\
    Sortino Ratio & 0.52   & 0.62   & 0.63 \\
    Max DD & 47.20\% & 41.70\% & 50.40\% \\
    10\% Worst DD & 21.10\% & 33.40\% & 24.80\% \\
    Return/maxDD & 0.2    & 0.3    & 0.2 \\
    Return/Worst 10\% DD & 0.5    & 0.4    & 0.4 \\
    Sampling & daily  & daily  & daily \\
    \bottomrule
    \end{tabular}%
  }%
\end{table}%

This trade-off between liquidity and performance remains a central barrier to building trustworthy PE replication products. Bridging this gap requires innovations that address both the volatility exposure and structural asymmetries in PE returns.

\subsection{Objective of This Study}

In response, this study proposes a replication methodology that leverages AI-powered graphical models and asymmetry-aware adjustments to mimic the stable, smoothed performance of quarterly PE benchmarks. The goal is to construct a liquid, scalable solution that meets institutional needs for transparency, capital efficiency, and resilience—while maintaining alignment with private equity’s superior long-term characteristics.

\subsection{Two-Step Approach}

Our methodology follows a two-step process designed to enhance both replication accuracy and performance robustness:

\begin{itemize}
    \item \textbf{Step 1: Decoding with Graphical Models.} We begin by estimating the composition of a liquid PE proxy (e.g., Stafford, TR, or Listed PE index) using graphical models. These probabilistic techniques generalize traditional regression and filtering approaches by modeling the dynamic relationships between asset weights and returns more accurately \cite{roncalli2007alternative, benhamou2024grip, Ohana2022deep}.

    \item \textbf{Step 2: Asymmetry Adjustment.} Next, we introduce a return transformation that reflects private equity’s downside smoothing behavior. Specifically, negative returns are reduced by an adjustment factor (e.g., 0.9), capturing the drawdown mitigation commonly observed in PE NAVs.
\end{itemize}

The transformation is defined as:
\[
R_t' = \begin{cases} R_t, & \text{if } R_t \geq 0, \\
AF \cdot R_t, & \text{if } R_t < 0, \end{cases}
\]
where \( AF = 0.9 \). This simple nonlinear mapping helps align the replicated performance with the asymmetric characteristics of traditional PE benchmarks.

To ensure model stability and scalability, additional constraints are introduced on asset weights based on historical boundaries. The overall process is summarized in Algorithm~\ref{alg:decoding_proxy}.

\begin{algorithm}
\caption{AI-Based Replication of Private Equity with Asymmetry Adjustment}
\label{alg:decoding_proxy}
\begin{algorithmic}[1]
\STATE Decode daily liquid PE proxy using equity index futures (Stafford, TR, or Listed PE)
\STATE Introduce asymmetry: scale down negative returns using adjustment factor (AF)
\STATE Train graphical model with maximum likelihood optimization
\STATE Conduct backtesting with continuous prediction correction
\STATE Apply sanity checks on historical weights
\STATE Enforce min/max constraints on weight evolution
\end{algorithmic}
\end{algorithm}

By integrating these steps, the methodology offers a trust-enhancing solution for replicating private equity in daily tradable form—balancing performance fidelity with liquidity and systemic reliability.

\begin{table}[htbp]
  \centering
  \caption{Performance of Bloomberg Private Equity Indexes\label{tab:bloomberg}}
  \resizebox{\columnwidth}{!}{%
    \begin{tabular}{llr}
    \toprule
           & \textbf{Bloomberg PEALL} & \textbf{Bloomberg PEBUY} \\
    \midrule
    Start Date & 31/03/2011 & 31/03/2011 \\
    End Date & 29/12/2023 & 29/12/2023 \\
    Annual Return & 11.4\% & 13.2\% \\
    Annual Volatility & 5.9\% & 7.6\% \\
    Skew   & -0.29  & -0.54 \\
    Kurtosis & 0.82   & 1.75 \\
    Sharpe Ratio & 1.95   & 1.73 \\
    Sortino Ratio & 4.07   & 2.34 \\
    Max DD & 5.2\% & 8.9\% \\
    10\% Worst DD & 2.\% & 2.1\% \\
    Return/maxDD & 2.2    & 1.5 \\
    Return/Worst 10\% DD & 5.6    & 6.2 \\
    Sampling & quarterly & quarterly \\
    \bottomrule
    \end{tabular}%
  }
\end{table}%

\subsection{Graphical Model Primer}\label{sec:Graphical model primer}

In building trust-enhancing frameworks for illiquid asset replication, it is essential to apply methodologies that combine transparency, adaptability, and interpretability. Graphical models serve as a powerful tool in this context. They offer a probabilistic framework for decoding daily return dynamics and asset weights in a way that accommodates nonlinearity and time-varying interactions—features characteristic of private equity performance.

Graphical models represent variables (such as asset weights or NAVs) as nodes in a network and dependencies as directed edges. This structure enables dynamic Bayesian inference, where latent states evolve over time and are continuously updated based on observed data. This is particularly important for creating replicable, transparent instruments that reflect the underlying mechanisms of private equity returns while being liquid and trustworthy.

To replicate the net asset value (NAV) of a private equity proxy, we define a model where asset weights evolve dynamically and determine the return at each time step. Suppose \( NAV_t \) is the observed net asset value at time \( t \), and \( r_t^{Eq}, r_t^{Fx}, r_t^{Ir}, r_t^{Co} \) are the returns of equity, foreign exchange, interest rate, and commodity assets. The predicted NAV at time \( t \), denoted \( \widehat{NAV}_t \), is given by:

\begin{equation}\label{eq:dynamic}
\widehat{NAV}_t = \widehat{NAV}_{t-1} \left( 1 + w^{Eq}_{t-1} r^{Eq}_t + \ldots + w^{Co}_{t-1} r^{Co}_t \right)
\end{equation}

This formulation highlights how portfolio dynamics are governed by prior weights and current returns. Unlike static regression approaches, graphical models account for dependencies across both time and asset types.

Key components of this approach include:

\begin{enumerate}
    \item \textbf{State Space Representation}: Asset weights are treated as latent variables evolving over time, influenced by past allocations and current market conditions. This state-space formulation supports dynamic modeling akin to, but more expressive than, Kalman filters.

    \item \textbf{Dynamic Inference}: Bayesian algorithms, including message passing, are used to estimate hidden states at each time step. This enables continuous recalibration of weights as new return observations arrive.

    \item \textbf{Cross-Asset Interaction}: Unlike traditional filtering methods, graphical models capture interdependencies across asset classes (e.g., FX volatility affecting equity allocations), reflecting the interconnected nature of market risks.

    \item \textbf{Self-Correcting Mechanisms}: The framework supports ongoing updates to the probability distribution of weights, making it highly adaptive and minimizing drift, which is crucial for institutional trust.
\end{enumerate}

Graphical models offer an interpretable and scalable method for decoding PE proxy behavior, providing a significant improvement over traditional linear regressions. Though underutilized in finance, they are foundational in machine learning and signal processing \cite{murphy2012machine}, and have powered early natural language and speech recognition systems (e.g., Siri).

\begin{figure}[ht]
    \centering
  \resizebox{\columnwidth}{!}{%
    \begin{tikzpicture}
        % Define node styles
        \tikzstyle{latent} = [circle, fill=gray!30, minimum size=1cm, draw=black, thick]
        \tikzstyle{observed} = [circle, fill=white, minimum size=1cm, draw=black, thick]
        \tikzstyle{legendbox} = [rectangle, draw=black, rounded corners, thick, fill=white]

        % Time axis
        \draw[thick, ->] (-1, 0) -- (7, 0) node[right] {Time};

        % Time steps
        \foreach \x\y in {0/1, 2/2, 4/3, 6/4}{
            \node at (\x, 0.5) {$t=\y$};
        }

        % Observed nodes (NAV) at each time step
        \foreach \x\y in {0/1, 2/2, 4/3, 6/4} {
            \node[observed] (NAV\x) at (\x, -1) {NAV$_\y$};
        }

        % Latent nodes for each category
        \foreach \x\y in {0/1, 2/2, 4/3, 6/4} {
            % Equity node
            \node[latent] (Eq\x) at (\x, -3) {Eq$_\y$};

            % FX node
            \node[latent] (Fx\x) at (\x, -4.5) {Fx$_\y$};

            % Interest Rate node
            \node[latent] (Ir\x) at (\x, -6) {Ir$_\y$};

            % Commodities node
            \node[latent] (Co\x) at (\x, -7.5) {Co$_\y$};
        }

        % Connections between observed nodes
        \foreach \x/\y in {0/2, 2/4, 4/6} {
            \draw[->, thick] (NAV\x) -- (NAV\y);
        }

        % Connections between latent nodes across time steps
        \foreach \x/\y in {0/2, 2/4, 4/6} {
            \draw[->, thick] (Eq\x) -- (Eq\y);
            \draw[->, thick] (Eq\x) -- (Fx\y);
            \draw[->, thick] (Eq\x) -- (Ir\y);
            \draw[->, thick] (Eq\x) -- (Co\y);

            \draw[->, thick] (Fx\x) -- (Eq\y);
            \draw[->, thick] (Fx\x) -- (Fx\y);
            \draw[->, thick] (Fx\x) -- (Ir\y);
            \draw[->, thick] (Fx\x) -- (Co\y);

            \draw[->, thick] (Ir\x) -- (Eq\y);
            \draw[->, thick] (Ir\x) -- (Fx\y);
            \draw[->, thick] (Ir\x) -- (Ir\y);
            \draw[->, thick] (Ir\x) -- (Co\y);

            \draw[->, thick] (Co\x) -- (Eq\y);
            \draw[->, thick] (Co\x) -- (Fx\y);
            \draw[->, thick] (Co\x) -- (Ir\y);
            \draw[->, thick] (Co\x) -- (Co\y);
        }

        % Connections from latent nodes to observed nodes
        \foreach \x in {0, 2, 4, 6} {
            \draw[<-, thick] (Eq\x) -- (NAV\x);
            \draw[<->, thin] (Fx\x) -- (Eq\x);
            \draw[<->, thin] (Ir\x) -- (Fx\x);
            \draw[<->, thin] (Co\x) -- (Ir\x);
        }

        % Add legend below
        \node[legendbox] (legend) at (3, -9) {
            \begin{tabular}{ll}
            \tikz\draw[latent] (0,0) circle (5pt); & Latent node (Eq, Fx, Ir, Co) \\
            \tikz\draw[observed] (0,0) circle (5pt); & Observed node (NAV$_t$) \\
            \end{tabular}
        };

        % Add title
        \node at (4, 1) {\textbf{Simplified Graphical Model for Time Inference}};
    \end{tikzpicture}
    }
    \caption{Simplified graphical model showing the relationship between observed NAv and inferred allocation as time goes by. For illustration purpose, we use different assets, with one being an Equity shortened in Eq, a second one an exchange rate shorted in Fx, a third one, an interest rates instrument shortened in Ir and finally a commodity asset shortened in Co.}
    \label{fig:graphical_model}
\end{figure}
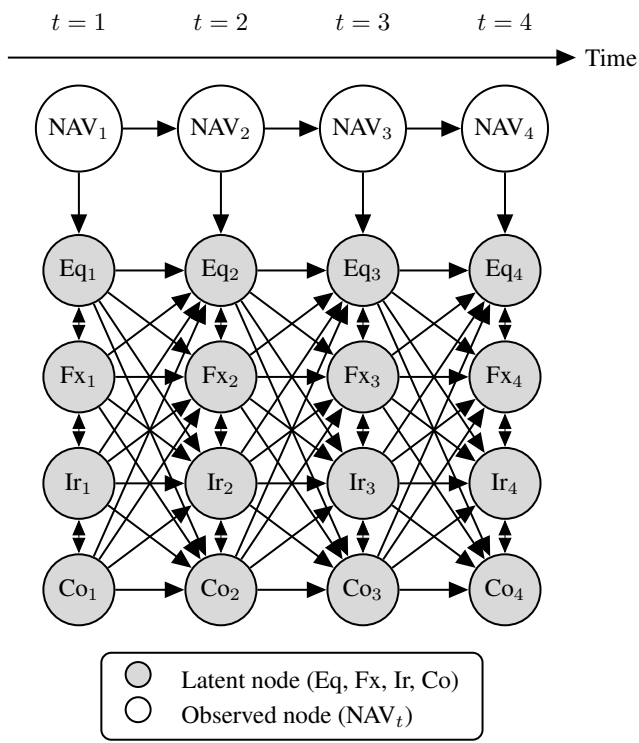

Figure~\ref{fig:graphical_model} illustrates a simplified version of the graphical model architecture. Each time step includes observed NAVs and latent weights across different asset categories. The model dynamically updates these weights while maintaining consistency over time, enabling robust real-time inference for daily liquid replication strategies.

\subsection{Asymmetric Factors}

One of the defining features of private equity benchmarks is their asymmetric return profile: strong upside potential paired with subdued downside volatility. This smoothing effect—attributed to infrequent valuation updates, discretionary pricing, and long investment horizons—presents a key challenge for liquid replication. To build trust in such proxies, we introduce asymmetry-aware components that capture these nonlinear characteristics, enhancing the realism and institutional viability of daily-traded replication strategies.

We incorporate two types of asymmetric overlays to emulate this behavior:

\begin{enumerate}
    \item \textbf{Tail Risk Overlay Using Volatility Instruments:} To mitigate drawdowns, we integrate a tail risk strategy based on volatility futures—specifically, dynamic long exposure to short-term (ST VIX) and medium-term (MT VIX) VIX futures. A machine learning framework identifies asymmetric risk environments by scanning three indicators:
    \begin{enumerate}
        \item The 20-day Volatility-Adjusted Return of the VIX Future,
        \item The VIX Curve Ratio (forward vs. spot), and
        \item The level of the VIX index, signaling deviations from historical norms.
    \end{enumerate}
    When the signals align, the model allocates capital across ST and MT VIX futures to hedge against market stress.

    Empirically, this approach boosts Sharpe ratios by over 70\% and improves drawdown-to-return ratios by a factor of 2.5 when applied to equity portfolios over the 2007–2024 period. Annual returns improve from 9\% to 12\%, while drawdowns are halved. This illustrates the benefit of explicitly embedding tail protection into a liquid replication framework.

    \item \textbf{Momentum with a Risk-Off Filter:} To complement tail risk hedging, we implement a momentum-based overlay inspired by CTA trend-following models. A hysteresis-based filter identifies prolonged negative equity trends and activates cross-asset long positions (e.g., in bonds, FX, or commodities) that typically exhibit negative correlation with equities.

    Backtested over 2010–2024, this risk-off momentum strategy displays a -36\% correlation with the S\&P 500 and delivers positive returns in 88\% of months when equities fall more than 5\%. It achieves an average monthly gain of 3.6\% during such drawdowns—outperforming traditional CTA benchmarks by a factor of two.
\end{enumerate}

Together, these asymmetric overlays serve dual purposes: they reduce maximum drawdowns and increase the fidelity of the replication relative to the smoothed performance of quarterly PE benchmarks. Most importantly, they help establish trust in daily liquid proxies by reducing vulnerability to shocks and creating a more stable, resilient investment experience—thereby advancing the broader goal of safe, transparent asset construction in illiquid markets.

\section{Results and Statistics\label{sec:Results and statistics}}

\subsection{Data Description}
To assess the validity and scalability of AI-driven private equity replication, we construct a comprehensive dataset of liquid proxies and industry benchmarks. Three daily liquid indexes are used as replication targets: the Erik Stafford proxy (Bloomberg: SHPEI Index), the Thomson Reuters Refinitiv Private Equity Benchmark (Bloomberg: TRPEI Index), and the S\&P Listed Private Equity Index (Bloomberg: SPLPEQNT Index). These represent real-time investable proxies for PE-like exposures.

The training period spans from 2005 to the end of 2010, while the out-of-sample test period begins in March 2011 to align with the availability of quarterly private equity benchmarks, including the Cambridge Associates, Preqin, Bloomberg Private Equity Buyout (PEBUY), and Bloomberg Private Equity All (PEALL) indexes. These quarterly benchmarks serve as performance anchors, reflecting the core return and risk characteristics institutional investors aim to replicate.

\subsection{Replication Accuracy and Trust Metrics}
To evaluate the effectiveness of AI-powered replication in building a transparent, high-frequency proxy for illiquid private equity exposure, we examine return characteristics, risk-adjusted performance, and benchmark alignment. Table~\ref{tab:decoding stats} presents the performance of the three decoded strategies. All exhibit robust annualized returns (17.1\%--17.7\%) with volatility levels significantly below their original benchmarks. Sharpe ratios exceed 1.2 across the board—demonstrating considerable improvement over raw liquid proxies (Table~\ref{tab:daily_index})—while preserving the essential return premia observed in traditional PE indexes.

\begin{table}[htbp]
  \centering
  \caption{Performance of AI-Decoded Strategies: Listed PE, TR, and Stafford \label{tab:decoding stats} }
  \resizebox{\columnwidth}{!}{%
    \begin{tabular}{lrrr}
    \toprule
           & \textbf{Listed PE} & \textbf{TR} & \textbf{Stafford}  \\
    \midrule
    Annual Return & 17.1\% & 17.7\% & 17.4\% \\
    Annual Volatility & 13.6\% & 14.0\% & 14.2\% \\
    Sharpe Ratio & 1.26 & 1.27 & 1.23 \\
    Max Drawdown & 17.6\% & 19.2\% & 19.2\% \\
    Return / Max DD & 1.0 & 0.9 & 0.9 \\
    \bottomrule
    \end{tabular}%
  }
\end{table}

The decoding approach captures the core performance patterns while reducing tail risks—an essential element for trust in liquid replication. Table~\ref{tab:correlation with benchmarks} confirms strong fidelity with benchmark trends. All strategies exceed 89\% correlation on a one-year horizon and maintain long-run correlations above 63\%. This validates the method’s capacity to deliver persistent exposure with high transparency.

\begin{table}[htbp]
  \centering
  \caption{Correlation with Original Benchmarks\label{tab:correlation with benchmarks}}
  \resizebox{\columnwidth}{!}{%
    \begin{tabular}{lrrrrrr}
    \toprule
    \textbf{Strategy} & \textbf{Lifetime} & \textbf{1Y} & \textbf{3Y} & \textbf{5Y} & \textbf{7Y} & \textbf{10Y} \\
    \midrule
    Stafford & 64\% & 94\% & 68\% & 81\% & 73\% & 71\% \\
    Listed PE & 63\% & 89\% & 65\% & 79\% & 75\% & 72\% \\
    TR & 69\% & 97\% & 71\% & 81\% & 78\% & 76\% \\
    \bottomrule
    \end{tabular}%
  }
\end{table}

\subsection{Cross-Strategy Consistency and Robustness}
The three decoding strategies exhibit strong internal coherence (Table~\ref{tab:correlation_between_strategies}), indicating methodological stability across diverse input benchmarks. This supports their scalability and reliability in institutional settings.

\begin{table}[htbp]
  \centering
  \caption{Cross-Correlation Among Decoded Strategies\label{tab:correlation_between_strategies}}
  \resizebox{0.7\columnwidth}{!}{%
    \begin{tabular}{|c|rrr}
    \toprule
           & \textbf{Stafford} & \textbf{Listed PE} & \textbf{TR} \\
    \midrule
    Stafford & 100\% & 83\% & 74\% \\
    Listed PE & 83\% & 100\% & 83\% \\
    TR & 74\% & 83\% & 100\% \\
    \bottomrule
    \end{tabular}%
  }
\end{table}

\subsection{Downside Mitigation and Temporal Consistency}
\begin{figure}[htbp]
  \centering
  \resizebox{\columnwidth}{!}{%
    \includegraphics{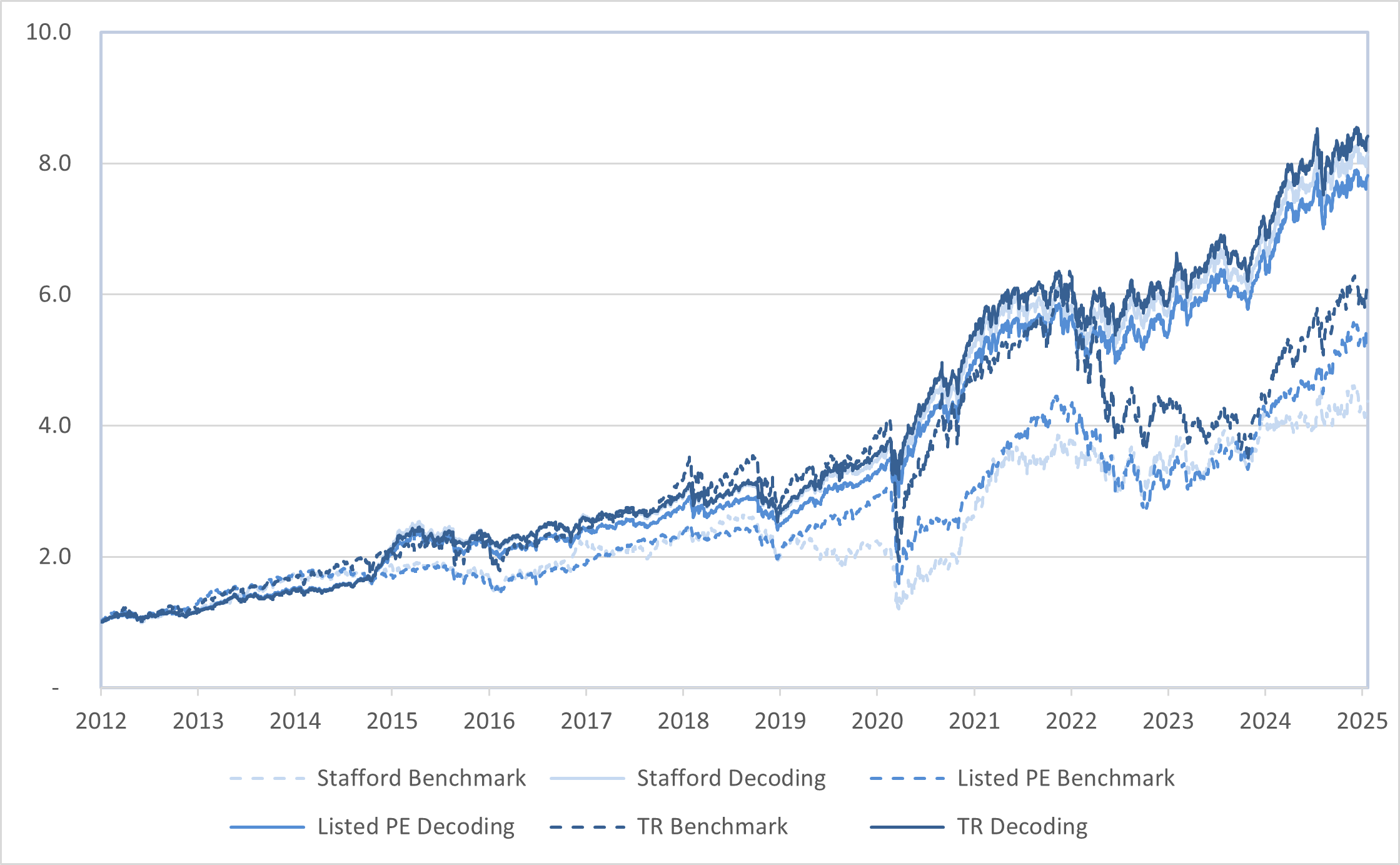}%
  }
  \caption{Comparison of AI-Decoded Strategies and Benchmarks\label{fig:pe_comparison}}
\end{figure}

Table~\ref{tab:yearly returns} and Figure~\ref{fig:pe_comparison} display annual returns versus benchmarks, illustrating how the AI-replicated strategies maintain performance in strong equity years and reduce losses in crisis periods (e.g., 2022, 2018). These features enhance institutional confidence in the strategy’s robustness and capital preservation.

\begin{table}[htbp]
  \centering
  \caption{Yearly returns  \label{tab:yearly returns}}
  \setlength{\tabcolsep}{3pt} % Adjust the column spacing (default is 6pt)
  \resizebox{1 \columnwidth}{!}{%
    \begin{tabular}{crrrrrr}
    \toprule
           & \multicolumn{3}{c}{Benchmark} & \multicolumn{3}{c}{Decoding} \\
    \multicolumn{1}{l}{\textbf{Years}} & \multicolumn{1}{c}{\textbf{Listed PE}} & \multicolumn{1}{c}{\textbf{TR}} & \multicolumn{1}{c}{\textbf{Stafford}} & \multicolumn{1}{c}{\textbf{Listed PE}} & \multicolumn{1}{c}{\textbf{TR}} & \multicolumn{1}{c}{\textbf{Stafford}} \\
    \midrule
    \midrule
    \textbf{2025} & 4.5\%  & 4.6\%  & 3.2\%  & 2.2\%  & 1.8\%  & 1.6\% \\
    \textbf{2024} & 24.0\% & 31.3\% & 2.7\%  & 16.0\% & 16.1\% & 16.2\% \\
    \textbf{2023} & 39.0\% & 4.4\%  & 23.1\% & 22.5\% & 21.3\% & 20.6\% \\
    \textbf{2022} & \textcolor[rgb]{ 1,  0,  0}{-29.0\%} & \textcolor[rgb]{ 1,  0,  0}{-31.1\%} & \textcolor[rgb]{ 1,  0,  0}{-11.1\%} & \textcolor[rgb]{ 1,  0,  0}{-4.2\%} & \textcolor[rgb]{ 1,  0,  0}{-4.0\%} & \textcolor[rgb]{ 1,  0,  0}{-1.1\%} \\
    \textbf{2021} & 41.8\% & 29.8\% & 43.4\% & 13.2\% & 12.8\% & 10.3\% \\
    \textbf{2020} & 4.5\%  & 25.6\% & 17.1\% & 52.3\% & 53.2\% & 52.2\% \\
    \textbf{2019} & 44.6\% & 37.4\% & 12.6\% & 31.6\% & 32.0\% & 30.3\% \\
    \textbf{2018} & \textcolor[rgb]{ 1,  0,  0}{-14.0\%} & \textcolor[rgb]{ 1,  0,  0}{-11.9\%} & \textcolor[rgb]{ 1,  0,  0}{-14.7\%} & \textcolor[rgb]{ 1,  0,  0}{-8.9\%} & \textcolor[rgb]{ 1,  0,  0}{-7.9\%} & \textcolor[rgb]{ 1,  0,  0}{-8.4\%} \\
    \textbf{2017} & 24.3\% & 31.5\% & 7.5\%  & 13.8\% & 14.3\% & 12.2\% \\
    \textbf{2016} & 13.6\% & 8.6\%  & 28.5\% & 9.3\%  & 10.2\% & 11.2\% \\
    \textbf{2015} & \textcolor[rgb]{ 1,  0,  0}{-3.2\%} & 6.7\%  & \textcolor[rgb]{ 1,  0,  0}{-9.4\%} & 7.7\%  & 9.8\%  & 8.0\% \\
    \textbf{2014} & \textcolor[rgb]{ 1,  0,  0}{-1.6\%} & 20.6\% & 7.8\%  & 32.5\% & 39.6\% & 40.2\% \\
    \textbf{2013} & 35.3\% & 42.6\% & 47.0\% & 32.2\% & 30.7\% & 31.7\% \\
    \textbf{2012} & 29.0\% & 20.4\% & 17.9\% & 16.0\% & 15.4\% & 16.0\% \\
    \end{tabular}%
    }
\end{table}%

Together, these results reinforce the thesis that artificial intelligence, when combined with return asymmetry and robust dynamic modeling, can serve as a credible and scalable solution for replicating private equity. The consistency across time, benchmarks, and metrics builds the transparency and trust necessary for adoption in risk-aware and liquidity-constrained portfolios.

\section{Conclusion}\label{sec:Conclusion}

This study proposes a novel and scientifically grounded framework for replicating private equity (PE) fund performance in liquid markets, addressing a critical challenge at the intersection of institutional demand, market stability, and asset accessibility. Motivated by the structural illiquidity of traditional PE investments and the rising need for transparency and scalability, we develop an AI-powered approach that combines dynamic graphical models with asymmetric return conditioning.

Our empirical results demonstrate that this methodology not only improves alignment with established quarterly benchmarks—such as those from Cambridge Associates and Preqin—but also mitigates drawdowns and enhances Sharpe ratios relative to existing liquid proxies. The framework captures key characteristics of private equity returns, including their asymmetric risk profile and crisis resilience, while providing daily liquidity and scalability through the use of public market instruments.

By introducing interpretable, probabilistic inference via graphical models and systematically incorporating tail-risk and momentum overlays, this work contributes to the growing literature on asset class replication and offers a pathway toward democratizing access to PE-like exposures. In doing so, it bridges the gap between trust in illiquid markets and the operational realities of modern portfolio construction.

Future research may extend this framework to other illiquid asset classes, such as real estates, infrastructure or private credit, and examine its integration into macroprudential policy design. By facilitating the construction of synthetic, liquid analogues to traditionally illiquid investments, this work contributes to a broader conversation about how financial innovation can support trust, stability, and accessibility in modern capital markets.

\bibliographystyle{ijcai25}
\bibliography{main}

\end{document}